\documentclass[journal=jpcafh,manuscript=article]{achemso}

\usepackage[version=3]{mhchem} % Formula subscripts using \ce{}
\usepackage{gensymb}
\usepackage{siunitx}
\usepackage{amsmath}
\usepackage{enumerate}
\usepackage{textgreek}
\author{Parham Rezaee}
\affiliation[Sharif University of Technology]
{Department of Chemistry, Sharif University of Technology, P.O. Box 11365-9516, Tehran, Iran}
\author{Mohsen Tafazzoli}
\email{tafazzoli@sharif.edu}
\phone{+98 21 66005718}
\fax{+98 21 66012983}
\affiliation[Sharif University of Technology]
{Department of Chemistry, Sharif University of Technology, P.O. Box 11365-9516, Tehran, Iran}

\title
  {Modeling of Light Gases Purification and Carbon Dioxide Capture by 1B--3N and 1N--3B Defects}

\abbreviations{NMR, h--BN, MD, DFT}

\begin{document}

%%%%%%%%%%%%%%%%%%%%%%%%%%%%%%%%%%%%%%%%%%%%%%%%%%%%%%%%%%%%%%%%%%%%%
%% The "tocentry" environment can be used to create an entry for the
%% graphical table of contents. It is given here as some journals
%% require that it is printed as part of the abstract page. It will
%% be automatically moved as appropriate.
%%%%%%%%%%%%%%%%%%%%%%%%%%%%%%%%%%%%%%%%%%%%%%%%%%%%%%%%%%%%%%%%%%%%%
\begin{tocentry}
\centering
\includegraphics{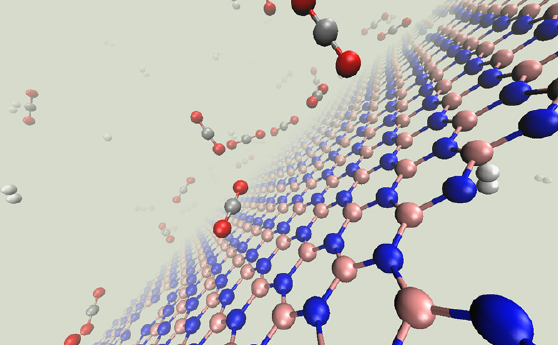}
\end{tocentry}

%%%%%%%%%%%%%%%%%%%%%%%%%%%%%%%%%%%%%%%%%%%%%%%%%%%%%%%%%%%%%%%%%%%%%
%% The abstract environment will automatically gobble the contents
%% if an abstract is not used by the target journal.
%%%%%%%%%%%%%%%%%%%%%%%%%%%%%%%%%%%%%%%%%%%%%%%%%%%%%%%%%%%%%%%%%%%%%
\begin{abstract}
In this study, we demonstrate that defected h--BN (1B--3N and 1N--3B defects) can be used as a suitable membrane for hydrogen purification and helium separation using density functional theory (DFT) calculations and molecular dynamics simulations (MD). At \SI{300}{\kelvin}, DFT calculations show that the selectivity of \ce{H2}/\ce{CO2}, \ce{H2}/\ce{N2}, \ce{H2}/\ce{CO}, and \ce{H2}/\ce{CH4} are \num{1d26}, \num{6d49}, \num{2d40}, and \num{1d94} respectively for 1B--3N, while they are \num{9d13}, \num{1d21}, \num{1d19}, and \num{1d42} for 1N--3B. Although selectivity of 1B--3N defect is much higher than 1N--3B defect, the permeance of this defect is much lower than industrial limits. To confirm the result obtained by DFT calculations for gas separation performance of 1N--3B defect, the classical molecular dynamics simulations were further carried out. Molecular dynamics simulations confirm the results of DFT calculations except \ce{H2}/\ce{CO2}. It demonstrates that the selectivity of \ce{H2} will grow, if temperature rises. This phenomenon is explainable by probability density distribution of the \ce{CO2} molecules at various temperatures. These simulations show that h--BN has good adsorption for \ce{CO2} molecules and is a suitable membrane for \ce{CO2} capture. Finally, the excellent selectivity along with acceptable permeance makes 1N--3B defects on h--BN, the promising membrane for He separation and \ce{H2} purification.
\end{abstract}

%%%%%%%%%%%%%%%%%%%%%%%%%%%%%%%%%%%%%%%%%%%%%%%%%%%%%%%%%%%%%%%%%%%%%
%% Start the main part of the manuscript here.
%%%%%%%%%%%%%%%%%%%%%%%%%%%%%%%%%%%%%%%%%%%%%%%%%%%%%%%%%%%%%%%%%%%%%
\section{Introduction}
The global tendency toward decreasing the greenhouse gases emissions, has resulted in increasing investment on hydrogen fuels \cite{dunn_hydrogen_2002}. One of the industrial processes that produces hydrogen gas is referred to as steam-methane reforming\cite{andrews_re-envisioning_2012}. As the initial step of this process, methane and excess steam react together and produce carbon monoxide and hydrogen gas at near \SI{800}{\degreeCelsius}. In second step, carbon monoxide and water react and produce carbon dioxide and additional hydrogen gas\cite{ockwig_membranes_2007}.
\begin{enumerate}[(I)]
    \item \ce{CH4 + H2O -> CO + 3H2}
    \item \ce{CO + H2O -> CO2 + H2}
\end{enumerate}
Hydrogen purification process is very important because impurities such as \ce{CO}, \ce{CO2}, \ce{N2} or \ce{CH4} can induce the fuel cell catalyst when hydrogen is used in\cite{pena_new_1996}. Today, gas separation through a combination can be done with too many processes such as amine adsorption, low-temperature distillation, cryogenic separation, and membrane separation. Hydrogen purification with membrane is currently considered as the most successful way due to low energy consumption, lower investment cost, possibility for continuous operation and its facile operation. Membranes which are suitable for \ce{H2} purification, are developed since 2001\cite{ismail_review_2001}. The most important membranes using for \ce{H2} purification are metal organic frameworks, zeolite, polymeric, and carbon based membranes\cite{tao_tunable_2014}. Permeability is a property of membrane that has an inverse proportional to membrane thickness. Furthermore, a good membrane should have controllable pore size, stable structure, and efficient permeance as membrane performance is determined by selectivity and permeability\cite{rezaee2020modified}. Theoretical and experimental investigations indicate that membranes with single-atomic thickness exhibit high permeance and acceptable selectivity\cite{zhu_theoretical_2015}. Pristine perfect monolayer sheets are impermeable to gases as small as \ce{He} atom\cite{bunch_impermeable_2008}.

Reaction between boric acid and urea at high temperature produces hexagonal boron nitride (h--BN), a monolayer material that is called ``white graphene''\cite{zhi_large-scale_2009,nag_graphene_2010}. Among the methods which have been utilized for synthesizing BN nanosheets are diverse synthesis and chemical blowing\cite{wang_two-dimensional_2016}. Diverse synthesis method includes mechanical cleavage, ball milling, irradiation by high-energy electron beam\cite{warner_atomic_2010}, and the reaction of boric acid and urea. In chemical blowing, ammonia borane acts as a precursor to generate low cost BN nanosheets\cite{darvish_ganji_hydrogen_2017}. Producing defect is one way to construct pore on membrane. Fischbein and co-workers used a focused electron beam to produce pores on graphene sheets. Heavy ion bombardment is another method for punching membranes to create nanopores\cite{jin_fabrication_2009}. Kotakoski and co-workers created boron nitride nanosheets with monovacancy defect using transmission electron microscopy (TEM) electron beam irradiation \cite{kotakoski_electron_2010}. Usually, the kinetic energy of the electrons in the TEM system quickly decays (in 10-100 \si{\femto\second}) in solid by electron-electron interaction, which is much faster than the structural changes (at least more than 1 \si{\pico\second}). Since the energy (120 \si{\kilo\electronvolt}) of the electron beam employed in the experiment of the TEM measurements is much larger than the threshold beam energies for knock on of both B (\SI{74}{\kilo\electronvolt}) and N (\SI{84}{\kilo\electronvolt}) atoms, both nitrogen and boron vacancies should appear at the same time without prominence\cite{yin_triangle_2010}. Using controlled energetic electron irradiation through a layer-by-layer sputtering process fabricates a single layer of h--BN. Furthermore, irregularity corrected high-resolution TEM resolves atomic defects with triangle shapes\cite{ryu_atomic-scale_2015}. More recently, Gilbert and co-workers fabricated triangular pores with desired sizes and diameters ranging from subnanometer to 6\si{\nano\meter} over a large area of h--BN sheet by conventional TEM technique. They demonstrated that the rate of pore growth can be controlled independently\cite{gilbert_fabrication_2017}, through precise control of the length of electron beam exposure.

In this work, we use the density functional theory (DFT) and molecular dynamics (MD) simulations to study the performance of defected hexagonal boron nitride on \ce{H2} purification and \ce{He} separation. Firstly, the energy barriers of gas permeating through two defects (1B--3N and 1N--3B) of h--BN are calculated using DFT to investigate the selectivity and permeability. Finally, the results of DFT calculations were confirmed by molecular dynamics simulations. In addition, these simulations demonstrated that \ce{CO2} molecules had the strongest adsorption on the surface in comparison with other species.

\section{Computational methods}
A large 2D nano sheet, \SI[product-units = power]{19.986 x 17.487}{\angstrom} in xy plane including 138 atoms of B and N is constructed to represent the 2D defective h--BN atomic layer. Geometry relaxation of defected hexagonal boron nitride were obtained by the Gaussian 09 program\cite{frisch_gaussian_2009} at the B3LYP/6-31G(d) level\cite{tian_expanded_2015}. On the basis of these optimized geometries, iso-electron density surfaces were plotted at iso-values \SI{0.0125}{\elementarycharge\angstrom^{-3}} to determine pore size of the defects\cite{chang_585_2017}. We used B3LYP/6-31G(d) calculations with DFT-D3 correction to find potential energy curves of a single \ce{He}, \ce{H2}, \ce{Ne}, \ce{CO2}, \ce{CO}, \ce{N2}, \ce{Ar}, and \ce{CH4} molecule crossing the pore center of defects\cite{zhu_theoretical_2017}. Furthermore, hirshfeld calculations were employed to analyze partial charges on atoms. Hirshfeld point charges more accurately reproduce the molecular electrostatic potentials than another methods such as natural population analysis (NPA)\cite{schrier_carbon_2012}.

To simulate gas separation using defected membranes, classical molecular dynamics simulations were performed with the LAMMPS\cite{LAMMPS} software package under constant particle number-volume-temperature (NVT) conditions using a Nose-Hoover thermostat set at wide range of temperature (250, 300, 350, 400, 450 and 500 \si{\kelvin}) and periodic boundary conditions were applied in all three dimensions\cite{yuan_molecular_2015}. The long-range coulomb interactions are computed by using the particle-particle particle-mesh algorithm (PPPM). The time step for the Newton's equations were integrated using 1 \si{\femto\second}\cite{tian_expanded_2015,nieszporek_alkane_2015}. To investigate potential effects of pressure, simulations were performed with various ratios (100:100, 200:200, 300:300, 400:400 and 500:500)\cite{alaghemandi_single_2015}.
The Dreiding force field which was validated by calculated potential energy curves for atoms in gas molecules and around pores in defected h--BN, is used to simulate membrane behavior. Two defected h--BN nanosheets consisting of 1984 boron and nitrogen atoms with an area in the xy plane of \SI{25.22}{\nano\meter\squared} with four defects on each ones, placed on $\pm 50$ z--axis.

The nonpolar \ce{CH4} and \ce{He} molecules are modeled as single spherical particles\cite{skoulidas_transport_2002}. Three--site model is used for \ce{H2} molecules in which quantum contribution is included through quadrupole interactions. This model has been successfully applied for hydrogen adsorption simulations in carbon nanostructures\cite{yang_molecular_2006}. The transferable potentials for phase equilibria-explicit hydrogen (ThaPPE-EH) model is used for \ce{N2} and \ce{CO2} molecules in the gas phase with an additional point charge site\cite{potoff_vaporliquid_2001}. In all cases the gas molecules are kept rigid with fixed bond lengths and angles. Interactions between these gases and the parameters of membrane's atoms are described by the Lorentz-Berthelot mixing rules.

\section{Results and discussion}
\begin{figure*}[t]
\centering
  \includegraphics[scale=0.8]{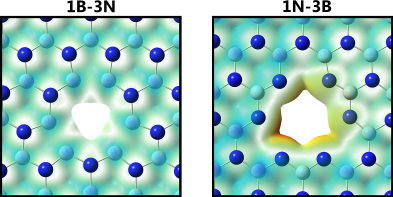}
  \caption{Top view of the fully optimized of 1B--3N and 1N--3B defects of h--BN. (B, pink; N, blue) (isovalue of \SI{0.0125}{\elementarycharge\angstrom^{-3}})}
  \label{fgr:opt}
\end{figure*}
The optimized structures (Figure \ref{fgr:opt}) show that pores size of 1B--3N and 1N--3B are analogous with kinetic diameter of \ce{He} (2.60\si{\angstrom}) \ce{H2} (2.89\si{\angstrom}), while smaller than those of other gas molecules. Therefore, these defects can be used as ideal pores on membrane for \ce{He} separation and \ce{H2} purification applications.
Next, we explore the \ce{He} and \ce{H2} selectivity of defected h--BN towards other gas molecules. The interaction energy between gas molecules and defected h--BN are computed by equation \ref{eq:e_int}:
\begin{equation}
\label{eq:e_int}
{E_{int}} = {E_{gas + sheet}} - {E_{gas}} - {E_{sheet}} 
\end{equation}
\begin{figure*}[b]
  \centering
  \includegraphics[scale=0.4]{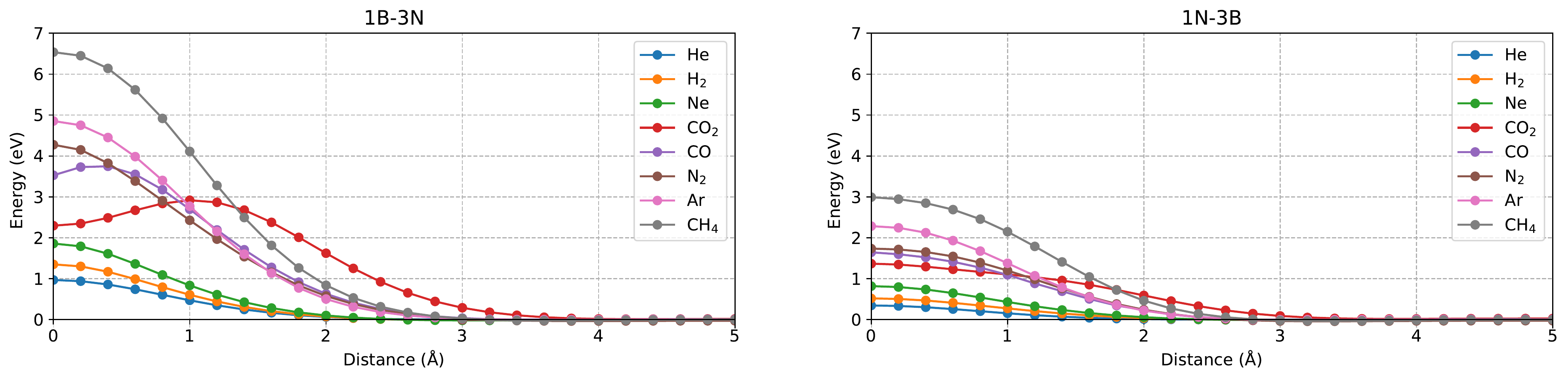}
   \caption{Potential energy curves for gases interacting with defected h--BN at the B3LYP/6-31G(d) level with DFT-D3 correction. Distance is from the center of mass of the gas molecule to the center of the pore.}
  \label{fgr:curve}
\end{figure*}
\begin{table*}[b]
\small
  \caption{Kinetic Diameter ($D_0$) of the Studied Gas Molecules, Adsorption Heights, Adsorption Energies and the Energy Barriers ($E_{barrier}$) for Gas Molecules Passing through Defects}
  \label{tbl:kindi}
  \begin{tabular*}{\textwidth}{@{\extracolsep{\fill}}lcccccccc}
    \hline
      &  \ce{He}  & \ce{Ne} & \ce{H2} & \ce{CO2} & \ce{Ar} & \ce{N2} & \ce{CO} & \ce{CH4}\\
    \hline
     ${D_0}(\si{\angstrom})$ & 2.60 &  2.75  & 2.89 & 3.30 & 3.40 & 3.64 & 3.76 & 3.80 \\
     ${H_{adsorption} (1B-3N)}(\si{\angstrom})$ & 2.61 & 2.69 & 2.65 & 3.15 & 2.91 & 3.20 & 3.19 & 3.22 \\
     ${H_{adsorption} (1N-3B)}(\si{\angstrom})$ & 2.58 & 2.67 & 2.61 & 3.11 & 2.91 & 3.18 & 3.17 & 3.21 \\
     ${E_{adsorption} (1B-3N)}(\si{\electronvolt})$ & -0.02 & -0.05 & -0.04 & -0.16 & -0.08 & -0.12 & -0.13 & -0.11 \\
     ${E_{adsorption} (1N-3B)}(\si{\electronvolt})$ & -0.02 & -0.04 & -0.04 & -0.11 & -0.06 & -0.09 & -0.11 & -0.10 \\
     ${E_{barrier} (1B-3N)}(\si{\electronvolt})$ &  1.144  & 1.992 & 1.349 & 2.904 & 4.858 & 4.310 & 3.750 & 6.952\\
    ${E_{barrier} (1N-3B)}(\si{\electronvolt})$ & 0.345 & 0.832 & 0.519 & 1.351 & 2.289 & 1.778 & 1.656 & 3.026\\
    \hline
  \end{tabular*}
\end{table*}
where $E_{gas+sheet}$ is the total energy of the gas molecule and defected h--BN, $E_{gas}$ is the energy of isolated gas molecule, and $E_{sheet}$ is the energy of defected h--BN. Potential energy curves for \ce{He}, \ce{Ne}, \ce{H2}, \ce{CO2}, \ce{CO}, \ce{N2}, \ce{Ar}, and \ce{CH4} interacting with defects of h--BN, such as 1B--3N and 1N--3B are mapped to obtain the barriers for gas to pass through (Figure \ref{fgr:curve}). The adsorption energies of \ce{H2}, \ce{He} and impurity gases are calculated to be in the range of $-0.02\sim-0.16$ \si{\electronvolt}, and the corresponding adsorption heights are in the range of $2.58\sim3.22$ \si{\angstrom} (see Table \ref{tbl:kindi}), revealing the physical adsorption character of all the gas molecules on the both 1B--3N and 1N--3B defects. The adsorption heights of impurity gases, are higher than that of \ce{He} and \ce{H2} molecules. It demonstrates that \ce{He} and \ce{H2} molecules are closer to the defects than other studied gas molecules. On the other hand, the adsorption energy of \ce{He} and \ce{H2} molecules are about -0.02 and -0.04 \si{\electronvolt} respectively for both defects, which are smaller than that of other gas molecules. Therefore, \ce{He} and \ce{H2} molecules are easier to desorb from the defects of h--BN and pass through it. The energy barrier for gas molecule passing through defects is defined as:
\begin{equation}
\label{eq:e_bar}
{E_{barrier}} = {E_{TS}} - {E_{SS}}
\end{equation}
\begin{figure*}[b]
\centering
  \includegraphics[scale=0.47]{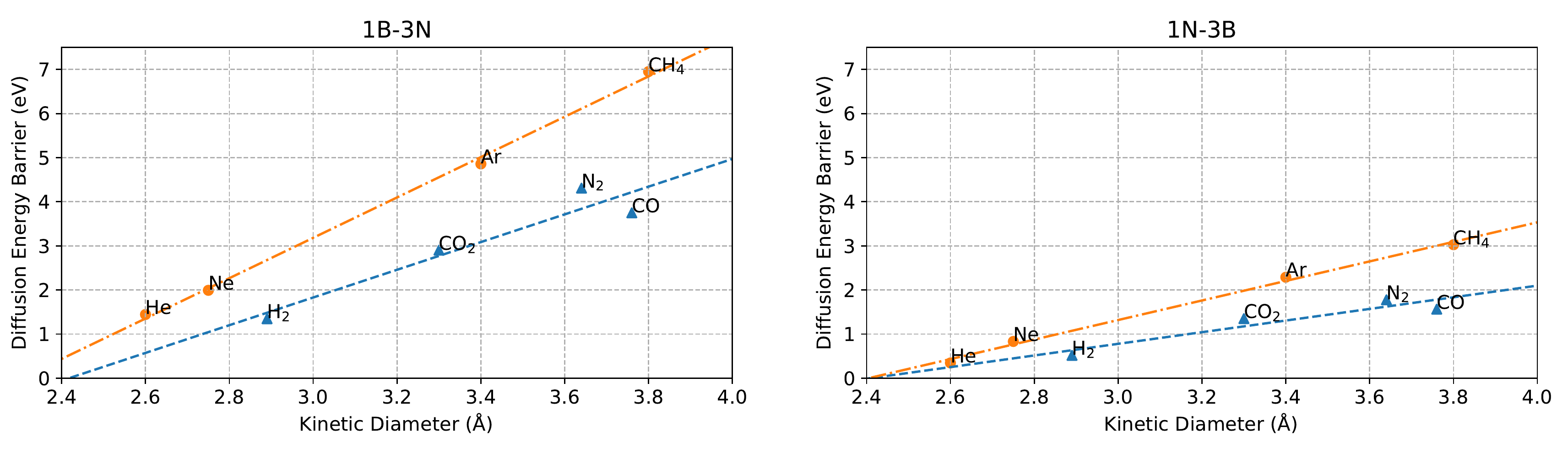}
  \caption{Dependence of the diffusion energy barriers on the kinetic diameters of different gas molecules}
  \label{fgr:kindi}
\end{figure*}
where $E_{TS}$ and $E_{SS}$ represent the interaction energies between gas molecules and defected h--BN in transition and steady state, respectively. The energy barriers of gas molecules passing through the defected h--BN are summarized in Table \ref{tbl:kindi}. Furthermore, this table represents the kinetic diameters of the studied gas molecules. Figure \ref{fgr:kindi} illustrates the dependence of diffusion energy barriers on the kinetic diameters\cite{B802426J} of the gas molecules. It can be seen that the diffusion energy barrier enhances with increasing the kinetic diameter of the gas. This phenomenon can be explained by increasing the electron density overlap between gas molecules and defects. Furthermore, we studied the electron overlaps between gas molecules and defects, as shown in Figure S1. Because there is no electron overlap between \ce{He} molecule and the defects, the energy barrier for \ce{He} molecule passing through the defects is the lowest. The electron overlap between \ce{CH4} molecule and center of defects makes the highest energy barrier among other molecules passing through defects. It is worth noting that the diffusion energy barriers of linear gas molecules (blue line) are lower than those of spherical gas molecules (orange line), even when the linear gas molecules have a larger kinetic diameter. It is owing to perpendicularly alignment of linear molecules to the membrane plane and making minimum energy barrier during the penetration process at transition state. In such configurations, the interactions between the gases and defects are thus minimized. This is in good agreement with their electron density overlaps with the h--BN defects.
Although kinetic diameter of \ce{H2} molecule is bigger than \ce{Ne} atom, the linear \ce{H2} molecule has a lower electron density overlap than a spherical \ce{Ne} atom. This can also be the case for \ce{N2} and \ce{CO} when compared with \ce{Ar}\cite{wang_two-dimensional_2016}.

The performance of membrane for \ce{He} separation and \ce{H2} purification is evaluated by the selectivity and permeance. To estimate the selectivity ($S_{\frac{X}{{gas}}}$) for X (\ce{He} and \ce{H2}) over other gas molecules on the basis of the calculated diffusion barrier energies, the Arrhenius equation was employed. The $S_{\frac{X}{{gas}}}$ between \ce{He} and \ce{H2} molecules over other gas molecules is defined as equation \ref{eq:sel}:
\begin{equation}
   \label{eq:sel}
   {S_{\frac{X}{{gas}}}} = \frac{{{r_X}}}{{{r_{gas}}}} = \frac{{{A_X}{e^{ - \frac{{{E_X}}}{{RT}}}}}}{{{A_{gas}}{e^{ - \frac{{{E_{gas}}}}{{RT}}}}}} 
\end{equation}
\begin{figure}[t]
\centering
  \includegraphics[scale=0.33]{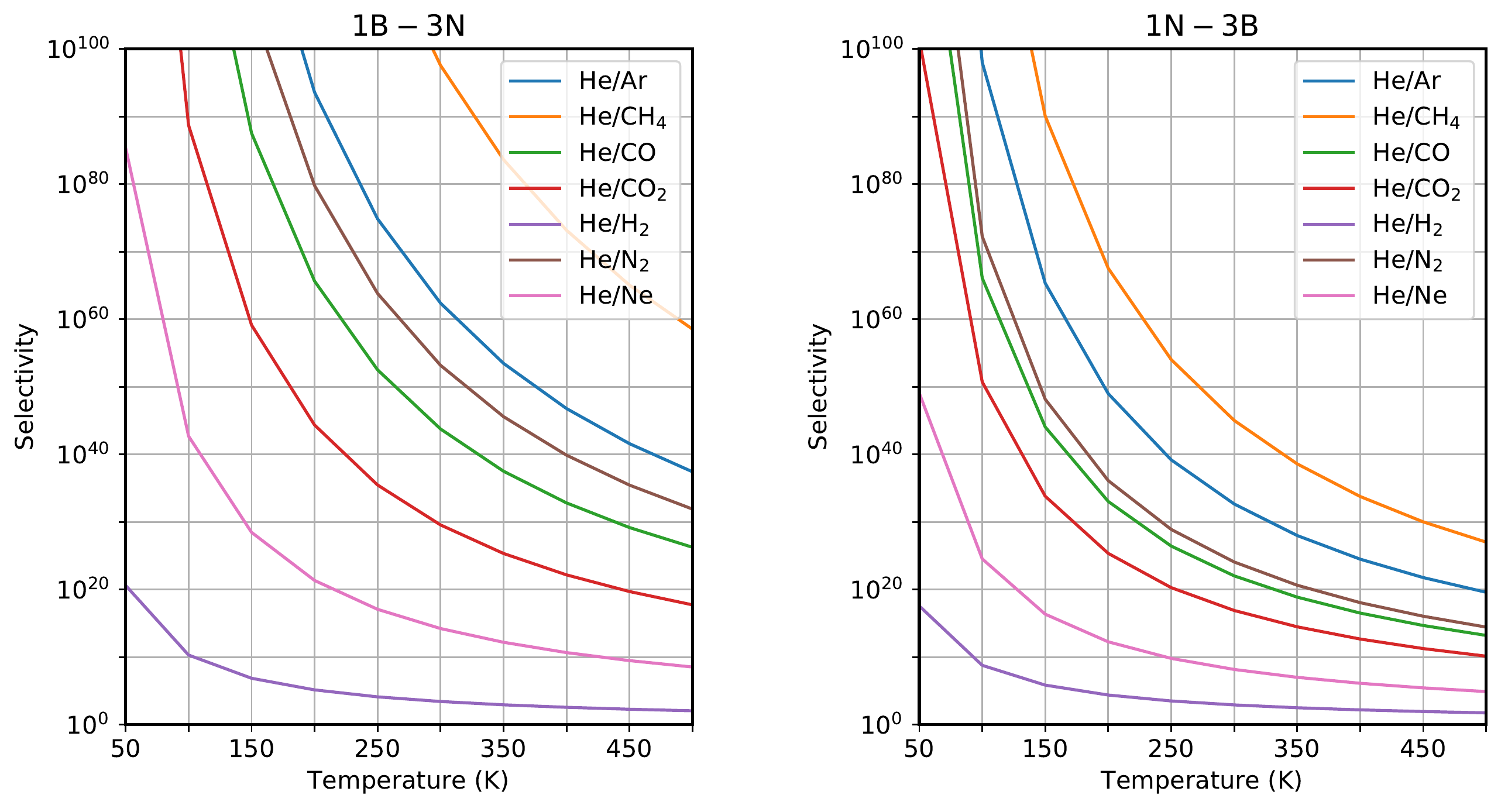}
  \caption{Selectivities of defected h--BN membrane for \ce{He} over other gases as a function of temperature.}
  \label{fgr:He}
\end{figure}
\begin{figure}[b]
\centering
  \includegraphics[scale=0.33]{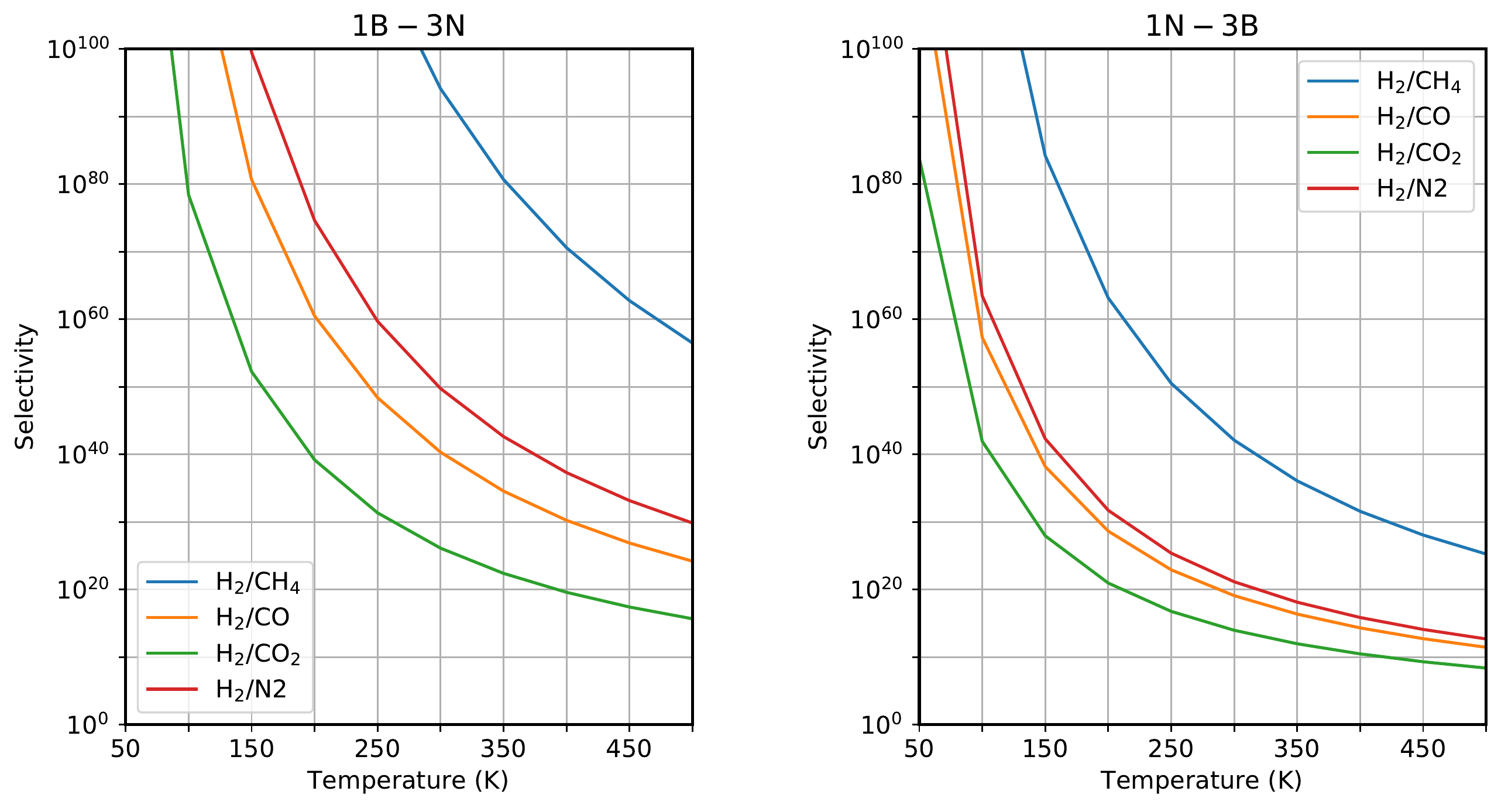}
  \caption{Selectivities of defected h--BN membrane for \ce{H2} over other gases as a function of temperature.}
  \label{fgr:H2}
\end{figure}
where r is the diffusion rate, A is the interaction (diffusion) prefactor, E is the diffusion barrier energy, R is the molar gas constant, and T is the temperature. Because A values are consistently around \SI{e11}{\per\second} for all gases\cite{gao_versatile_2017}, here we assume that the prefactors of all the gases are identical. The calculated selectivities for helium separation at different temperatures are shown in Figure \ref{fgr:He}. It can be seen that the selectivity decreases with increasing temperature. The selectivities for \ce{He} over all other gas molecules at room temperature (\SI{300}{\kelvin}) as well as those of some previously proposed porous monolayers are comprised in Table \ref{tbl:selhe}. The results indicate that 1B--3N defect has the highest \ce{He}/gas selectivities among other membranes. It is worth noting that 1N--3B defect shows good selectivities like the other membranes. This demonstrates that defected h--BN (both 1B--3N and 1N--3B) represents a very promising monolayer membrane for \ce{He} purification. On the other hand, most membranes are unable to separate \ce{He} and \ce{H2}, due to their nearly identical kinetic diameters. However, we found that defected h--BN can separate \ce{He} from \ce{H2} with a selectivity of about \num{3d4} for 1B--3N and \num{900} for 1N--3B at room temperature. In addition, hydrogen purification using these defects, will be done perfectly as shown in Figure \ref{fgr:H2}. At room temperature (\SI{300}{\kelvin}), the selectivities of \ce{H2}/\ce{CO2}, \ce{H2}/\ce{N2}, \ce{H2}/\ce{CO}, and \ce{H2}/\ce{CH4} are calculated to be \num{1d26}, \num{6d49}, \num{2d40}, and \num{1d94} for 1B--3N and \num{9d13}, \num{1d21}, \num{1d19}, and \num{1d42} for 1N--3B, respectively, which are much higher than those for most of other monolayer membranes (Table \ref{tbl:selh2}). Therefore, both defects are good choices for helium separation and hydrogen purification in the ideal condition.
\begin{table*}[t]
\small
  \caption{Comparison of the Monolayer Defected h--BN Selectivities for \ce{He} over other Gases in Room Temperature (\SI{300}{\kelvin}) with those of other Proposed Porous Membranes}
  \label{tbl:selhe}
\begin{tabular*}{\textwidth}{@{\extracolsep{\fill}}lccccccc}
    \hline
membrane & \ce{He}/\ce{Ne} & \ce{He}/\ce{H2} & \ce{He}/\ce{CO2} & \ce{He}/\ce{Ar} & \ce{He}/\ce{N2} & \ce{He}/\ce{CO} & \ce{He}/\ce{CH4} \\
    \hline
1B--3N (This work) & \num{2d14} & \num{3d4} & \num{4d29} & \num{2d62} & \num{2d53} & \num{6d43} & \num{4d97} \\
1N--3B (This work) & \num{2d8} & \num{9d2} & \num{8d16} & \num{5d32} & \num{1d24} & \num{1d22} & \num{1d45} \\
CTF-0\cite{wang_two-dimensional_2016} & \num{4d6} & \num{4d2} & \num{4d16} & \num{5d35} & \num{2d27} & \num{5d24} & \num{6d38} \\
Silicene\cite{hu_helium_2013} & \num{2d3} & --- & --- & \num{2d18} & --- & --- & --- \\
Polyphenylene\cite{blankenburg_porous_2010} & \num{6d2} & \num{9d2} & \num{6d15} & \num{1d30} & \num{2d22} & \num{4d20} & --- \\
Graphdiyne\cite{bartolomei_graphdiyne_2014} & \num{27} & --- & --- & --- & --- & --- & \num{1d24} \\
\ce{g-C3N4}\cite{li_efficient_2015} & \num{1d10} & \num{1d7} & --- & \num{1d51} & \num{1d34} & \num{1d30} & \num{1d65} \\
\ce{C2N7}\cite{zhu_c_2015} & \num{3d3} & --- & \num{8d18} & \num{4d18} & \num{3d12} & --- & \num{7d31} \\
    \hline
  \end{tabular*}
\end{table*}
\begin{table*}[t]
\small
  \caption{Comparison of the Monolayer Defected h--BN Selectivities for \ce{H2} over other Gases at Room Temperature (\SI{300}{\kelvin}) with those of other Proposed Porous Membranes}
  \label{tbl:selh2}
  \begin{tabular*}{\textwidth}{@{\extracolsep{\fill}}lcccc}
    \hline
    membrane & \ce{H2}/\ce{CO2} & \ce{H2}/\ce{N2} & \ce{H2}/\ce{CO} & \ce{H2}/\ce{CH4} \\
    \hline
    {1B-3N (This work)} & \num{1d26} & \num{6d49} & \num{2d40} & \num{1d94} \\
    {1N-3B (This work)} & \num{9d13} & \num{1d21} & \num{1d19} & \num{1d42} \\    
    {$\gamma$-GYH}\cite{sang_excellent_2017} & \num{9d17} & \num{1d26} & \num{7d23} & \num{2d49} \\
    {Polyphenylene\cite{blankenburg_porous_2010}} & \num{7d16} & \num{2d23} & \num{5d21} & --- \\
    {Phosphorene}\cite{zhang_hydrogen_2016} & \num{1d15} & \num{1d13} & \num{1d12} & \num{1d21} \\
    {Graphenylene}\cite{song_graphenylene_2013} & \num{1d14} & \num{1d13} & \num{1d12} & \num{1d34} \\
    {CTF-0}\cite{wang_two-dimensional_2016} & \num{9d13} & \num{4d24} & \num{1d22} & \num{2d36} \\
    {$\gamma$-GYN}\cite{sang_excellent_2017} & \num{2d13} & \num{2d21} & \num{1d18} & \num{2d46} \\
    {Rhombic-graphyne\cite{zhang_tunable_2012}} & --- & \num{1d19} & \num{1d16} & \num{1d41} \\
     {Silicene\cite{hu_helium_2013}} & \num{1d11} & \num{1d11} & \num{1d11} & \num{1d22} \\
    {Porous graphene\cite{jiang_porous_2009}} & --- & --- & --- & \num{1d22} \\
    {Germanene}\cite{chang_585_2017} & \num{1d10} & \num{1d14} & \num{1d18} & \num{1d36} \\
    {Fused pentagon network\cite{zhu_theoretical_2015}} & \num{1d7} & \num{1d8} & \num{1d7} & \num{1d31} \\
    \ce{g-C2O}\cite{zhu_theoretical_2017} & \num{3d3} & \num{2d6} & \num{2d5} & \num{4d23} \\
    \ce{g-C3N3\cite{ma_computational_2014}} & --- & \num{1d6} & \num{1d4} & \num{1d26} \\
    {Graphdiyne\cite{zhang_tunable_2012}} & --- & \num{1d3} & \num{1d3} & \num{1d10} \\
    {Silica\cite{de_vos_high-selectivity_1998}} & \num{10} & \num{1d2} & --- & \num{1d3}\\
    \hline
  \end{tabular*}
\end{table*}
\begin{figure*}[t]
\centering
  \includegraphics[scale=0.33]{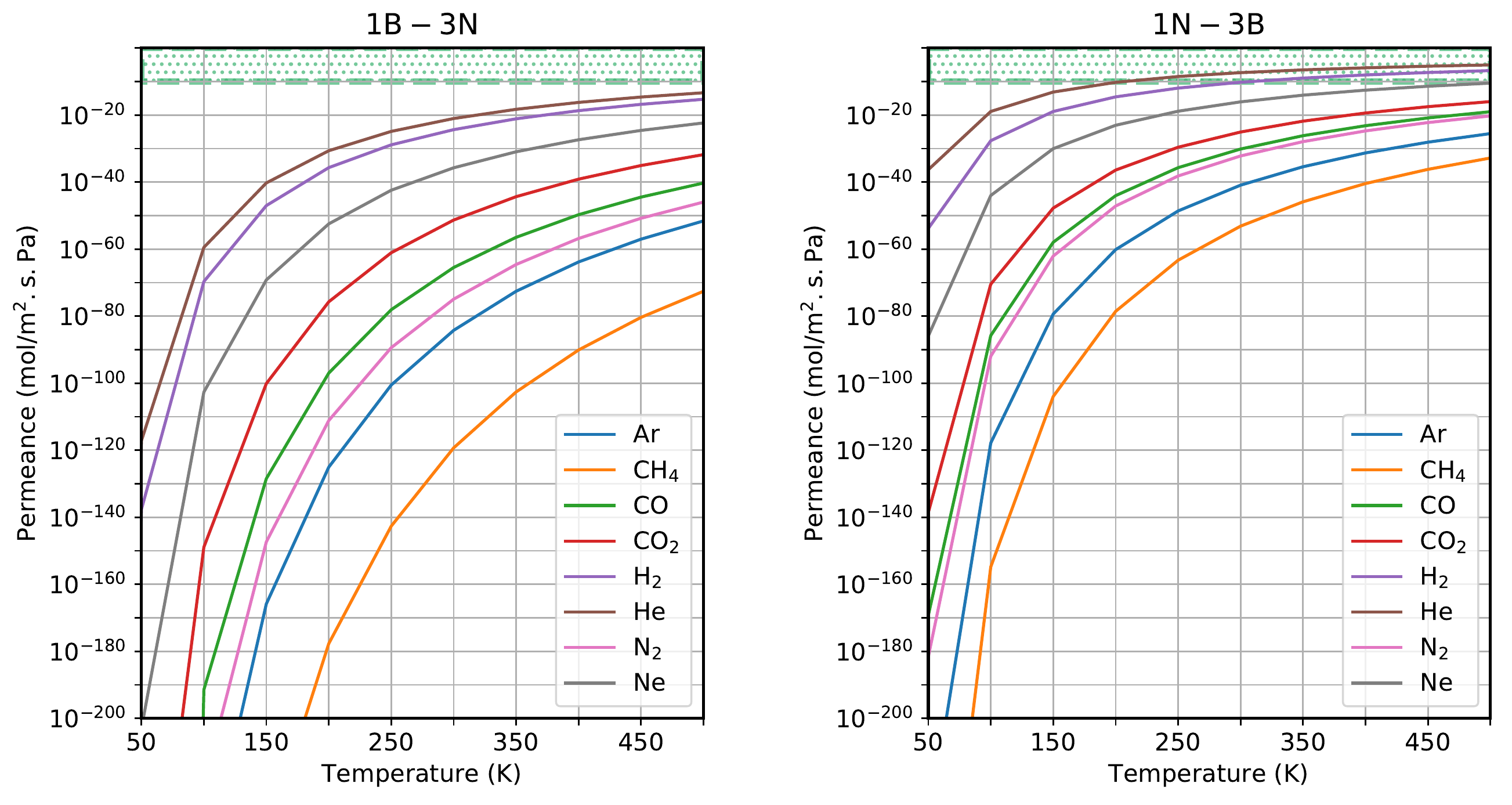}
  \caption{Permeances versus temperature for the studied gases passing through defected h-BN membranes as a function of temperature.}
  \label{fgr:per}
\end{figure*}

The performance of a membrane is characterized not only by the selectivity but also by the permeance. With the aid of the calculated energy barriers, we further used the kinetic theory of gases and the Maxwell-Boltzmann distribution to estimate the gas permeances. The number of gas particles colliding with the defected area of h--BN (N) can be expressed by:
\begin{equation}
N = \frac{P}{{\sqrt {2\pi MRT} }}
\end{equation}
P is the gas pressure, here it is taken as $\SI{3e5}{\pascal}$, M is the molar mass, R is the universal gas constant, and T is the gas temperature. The probability of a particle diffusing through the pore is given by \cite{rezaee_graphenylene1_2020}:
\begin{equation}
f = \int_{{v_B}}^\infty  f (v)dv
\end{equation}
where $v_B$ represents the velocity corresponding to the energy barrier and $f(v)$ is the Maxwell velocity distribution. Thus, the flux can be calculated as $F = N \times f$\cite{ji_heptazine-based_2016}. It should be noted that some factors that are difficult to consider in our theoretical approach such as the orientation of the non-monoatomic molecules, may lead to overestimation of flux. Assuming a pressure drop of $\Delta P$ equals $\SI{1e5}{\pascal}$. The calculated permeances for the studied gases, which reached by $p = F / \Delta P$ are shown in Figure \ref{fgr:per}, where the dashed line represents the industrially acceptable permeance limit for gas separation. For \ce{He}, the 1N--3B defect shows excellent performance in terms of its high permeance above \SI{300}{\kelvin}, exceeding the industrial standard by around 3 orders of magnitude at \SI{300}{\kelvin}. For \ce{H2}, the permeance is also higher than the industrially accepted standard, and exceeds the industrial standard by around 3 orders of magnitude at \SI{500}{\kelvin}. Although 1B--3N defect shows high selectivity, it has unacceptable permeance.

In order to further investigate the permeance and selectivity of the defected h--BN for the \ce{He}/\ce{H2}, \ce{H2}/\ce{CO2}, \ce{H2}/\ce{N2}, and \ce{H2}/\ce{CH4} binary gas mixtures, the permeation process was also modeled using the MD simulations. Figure S2--S6 shows the progress of separation of the equimolar modeling systems with ratios 100:100, 200:200, 300:300, 400:400 and 500:500 at 250,300, 350, 400, 450, and 500 \si{\kelvin} after the simulation time of \SI{5}{\nano\second}. Based on the final configurations of the modeling systems, we can obtain the numbers of gas molecules passing through the membranes after \SI{5}{\nano\second} by counting the number of molecules in the vacuum regions. Then, the actual selectivity (S) of gas A over gas B can be defined as:

In order to further investigate the permeance and selectivity of the defected h--BN for the \ce{He}/\ce{H2}, \ce{H2}/\ce{CO2}, \ce{H2}/\ce{N2}, and \ce{H2}/\ce{CH4} binary gas mixtures, the permeation process was also modeled using the MD simulations. Figure S2--S6 shows the progress of separation of the equimolar modeling systems with ratios 100:100, 200:200, 300:300, 400:400 and 500:500 at 250,300, 350, 400, 450, and 500 \si{\kelvin} after the simulation time of \SI{5}{\nano\second}. Based on the final configurations of the modeling systems, we can obtain the numbers of gas molecules passing through the membranes after \SI{5}{\nano\second} by counting the number of molecules in the vacuum regions. Then, the actual selectivity (S) of gas A over gas B can be defined as:
\begin{equation}
{S_{\frac{A}{B}}} = \frac{{\frac{{{x_A}}}{{{x_B}}}}}{{\frac{{{y_A}}}{{{y_B}}}}} = \frac{{\frac{{{N_A}}}{{{N_{0,A}}}}}}{{\frac{{{N_B}}}{{{N_{0,B}}}}}}
\end{equation}
\begin{figure*}
\centering
  \includegraphics[scale=0.64]{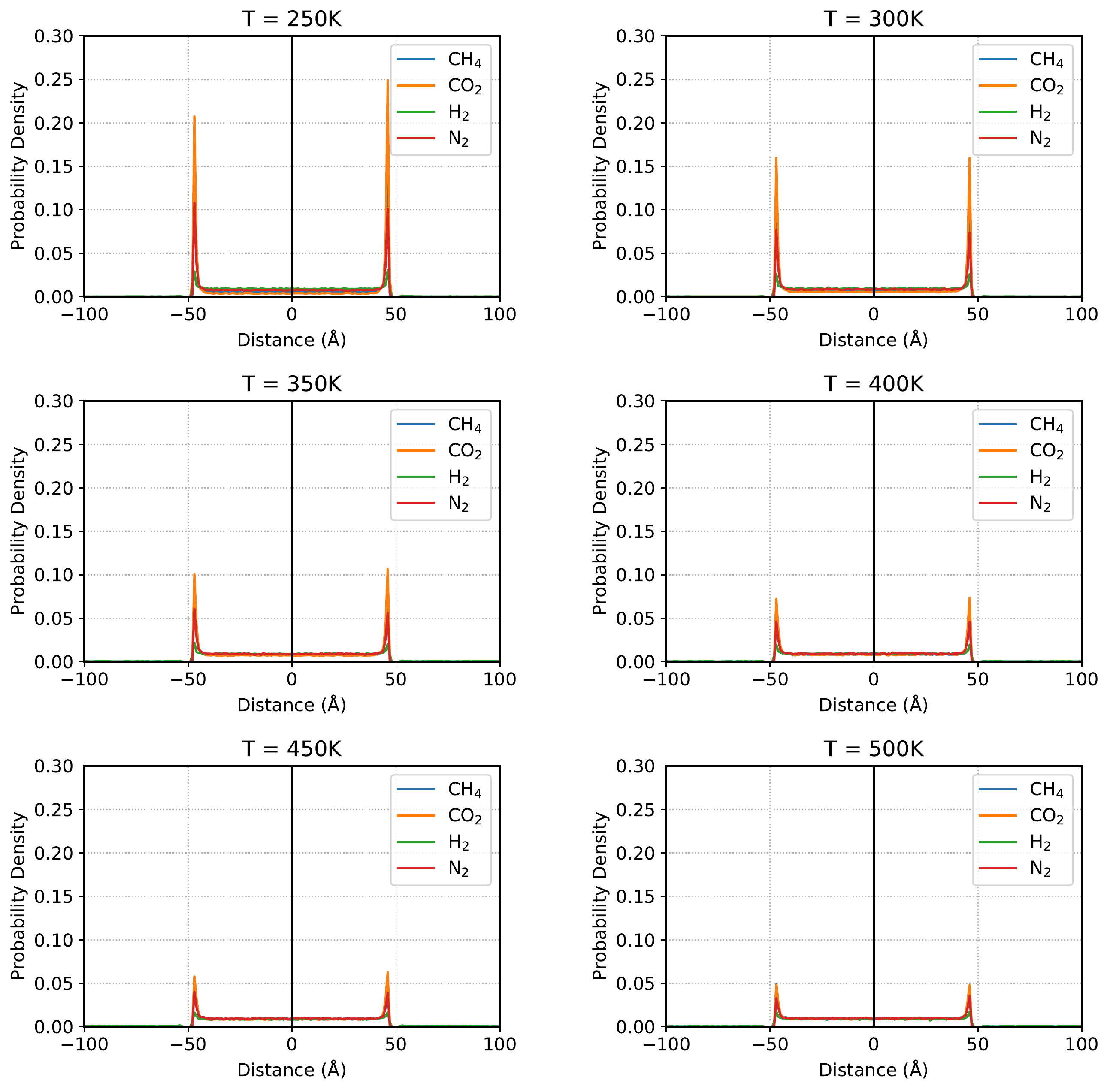}
  \caption{Probability density distribution of the molecules, as a function of distance to defected h--BN (1N--3B defects) plane, from the \SI{\pm 100}{\angstrom} simulation results for ratio 100:100.}
  \label{fgr:prob}
\end{figure*}
\begin{figure*}[t]
\centering
  \includegraphics[scale=0.4]{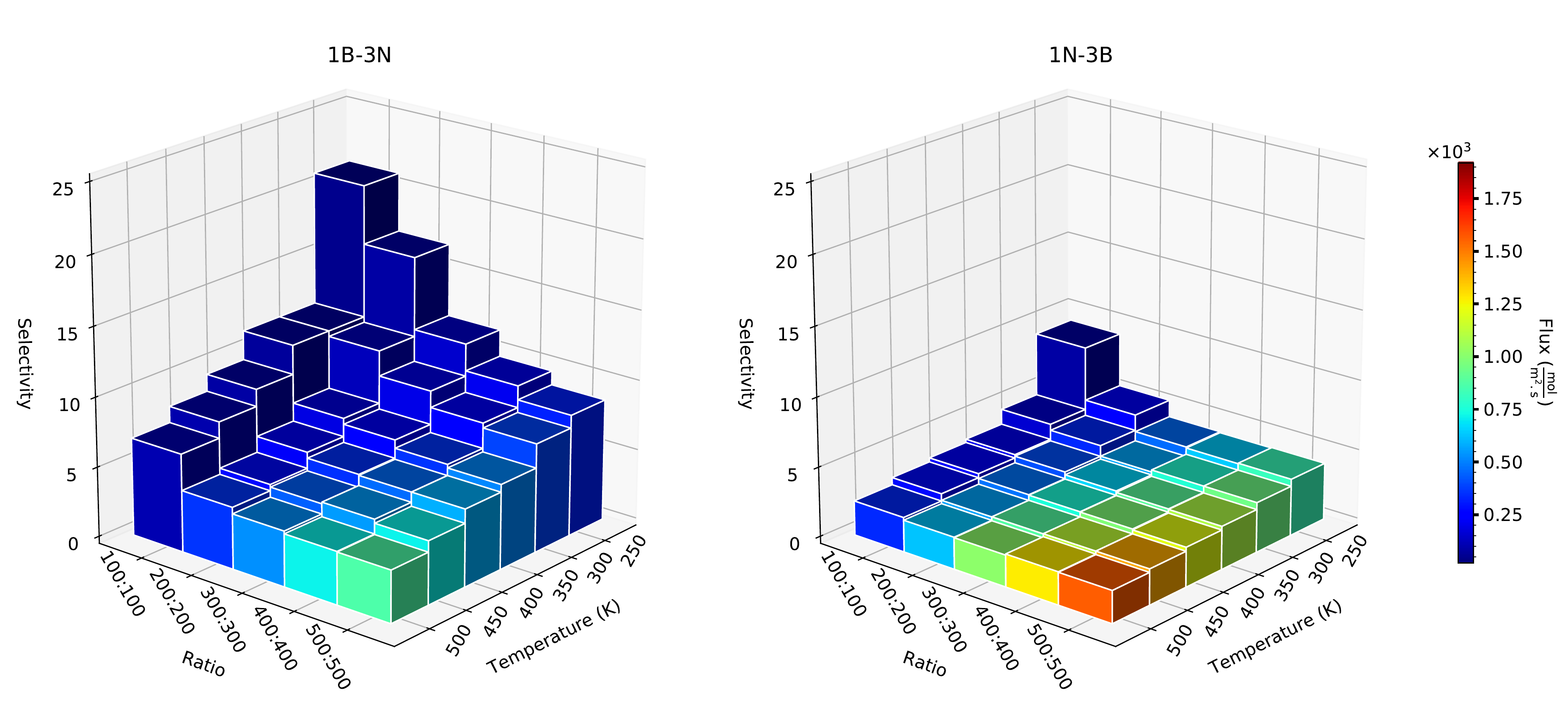}
  \caption{Comparison of \ce{H2} flux and \ce{He}/\ce{H2} ideal selectivity of 1B--3N and 1N--3B defects in different ratios and temperatures. The color bar shows the flux of \ce{H2} molecules in different situations.}
  \label{fgr:rstf}
\end{figure*}
where $x_A$ ($x_B$) and $y_A$ ($y_B$) are the mole fractions of component A (B) in the vacuum regions and gas reservoir, respectively, and $N_A$ ($N_B$) and $N_{0,A}$ ($N_{0,B}$) are the corresponding numbers of molecule A (B)\cite{wesolowski_pillared_2011}. It can be seen that the 1N--3B defect shows high transmissivities for \ce{He} and \ce{H2} gas molecules. Given that DFT studied showed good permeance for \ce{H2} gas molecules from 1N--3B defects, so MD simulation to purify \ce{H2} from \ce{CO2}, \ce{N2} and \ce{CH4} gas molecules is applied. On the other hand, there is no \ce{N2} and \ce{CH4} passing through the 1N--3B defect after \SI{5}{\nano\second} due to their large kinetic diameter, while \ce{CO2} have considerable transmissivities. Therefore, 1N--3B exhibits excellent selectivity and permeance for the \ce{H2}/\ce{N2} and \ce{H2}/\ce{CH4} mixtures, but the actual selectivity for \ce{H2}/\ce{CO2} is lower than the other gases in range of temperatures because it allows both \ce{CO2} and \ce{H2} molecules to pass through. In general, the selectivity of a membrane strongly depends on temperature and generally decreases with increasing temperature. It is due to the kinetic energy and the probability of gas molecules which collide to the membrane, will increase by raising temperature. So we can see that the number of \ce{He} and \ce{H2} molecules passing through the defect, will grow. On the other hands the number of \ce{CO2} molecules passing through the defect will decrease. It is because of nitrogen atoms in h--BN membrane, adsorbed \ce{CO2} molecules at low temperature and the probability of \ce{CO2} molecules passing through the defects will increase. Unfortunately, when temperature raises, the number of \ce{CO2} molecules which adsorb on the h--BN membrane will decrease. So the probability of \ce{CO2} molecules which being placed near the defects, will reduce.

Thus the number of \ce{CO2} molecules which passing through the defects will decrease. Tables S1--S3 present the selectivity and the flow of hydrogen purification in different ratios at different temperatures. The probability density distribution (Figure \ref{fgr:prob}) shows stronger adsorption of the \ce{CO2} molecules on the surface, and weaker adsorption for the other species at ratio 100:100. Figures S7--S10, also show the probability density distribution for ratios 200:200, 300:300, 400:400, and 500:500. Increasing the pressure decreases the probability density distribution near the surfaces. It means that the adsorption sites are limited and increasing the number of molecules, reduces the probability density distribution near the h--BN.

Moreover, the flow demonstrates the membrane permeability quantitatively as below:
\begin{equation}
F = \frac{N}{{A \times t}}
\end{equation}
where N shows the moles of permeated gas molecules through the membrane, A is the area of membrane and t is the time duration\cite{du_separation_2011}. Figure \ref{fgr:rstf}, illustrates the selectivity and the flow of helium separation from hydrogen at different ratios and temperatures. Figure \ref{fgr:rstf} reconfirms this fact, increasing temperature and pressure for the systems which the molecules have less adsorption on surface, like separation \ce{He} from \ce{H2}, decrease selectivity.

\section{Conclusions}
In this work, we theoretically investigated the helium separation and hydrogen purification performance of two defects of h--BN (1B--3N and 1N--3B) through DFT calculations. The results show that the energy barriers of gases (\ce{He}, \ce{H2}, \ce{Ne}, \ce{CO}, \ce{CO2}, \ce{N2}, \ce{Ar}, and \ce{CH4}) passing through these defects are related to their electron density overlaps. Diffusion energy barrier will be higher, if the electron density overlaps increase. The performance of defected h--BN for \ce{He} separation and \ce{H2} purification has been quantitatively investigated using the diffusion energy barrier. The selectivity of gas molecules decrease with raising temperature and the permeance of \ce{He} and \ce{H2} molecule is higher than the acceptable industrial standard for 1N--3B defect. Molecular dynamics simulations confirm the results of DFT calculations except the results of \ce{H2}/\ce{CO2}. It demonstrates that the selectivity of \ce{H2} will grow, if temperature raises. This phenomenon is explainable according to the probability density distribution of the \ce{CO2} molecules at different temperatures. In summary, the excellent selectivity along with acceptable permeance makes 1N--3B defects on h--BN, the promising membrane for \ce{He} separation and \ce{H2} purification.

\section{ASSOCIATED CONTENT}
\begin{suppinfo}
Iso-electron density surfaces of molecules passing through defects, number of permeated molecules versus simulation time through defects at different ratios and temperatures, comparison of \ce{H2} flow and \ce{H2}/\ce{CH4}, \ce{H2}/\ce{N2}, \ce{H2}/\ce{CO2} ideal selectivity of 1N--3B defects in different ratios and temperatures, probability density distribution of the molecules verses distance for ratios 200:200, 300:300, 400:400, and 500:500.
\end{suppinfo}

\begin{acknowledgement}
We would like to express our very great appreciation to Shahab Rezaee for his valuable contributions on computer codes used for analysis and visualizations during this research. Our special thanks go to the Department of Chemistry and High Performance Computing Centre (HPCC) of Sharif University of Technology for generously supplying the computer facilities.
\end{acknowledgement}

\providecommand{\latin}[1]{#1}
\makeatletter
\providecommand{\doi}
  {\begingroup\let\do\@makeother\dospecials
  \catcode`\{=1 \catcode`\}=2 \doi@aux}
\providecommand{\doi@aux}[1]{\endgroup\texttt{#1}}
\makeatother
\providecommand*\mcitethebibliography{\thebibliography}
\csname @ifundefined\endcsname{endmcitethebibliography}
  {\let\endmcitethebibliography\endthebibliography}{}

\end{document}